\documentclass[aps,prb,graphicx,twocolumn,superscriptaddress,showpacs]{revtex4-2}
\usepackage{amsmath,graphicx,latexsym,times,color}
\usepackage{setspace}
\usepackage[hidelinks,colorlinks=true, allcolors=blue]{hyperref}
\usepackage{array}
\usepackage{textcomp}
\usepackage{titlesec}
\usepackage{physics}
\usepackage{gensymb}
\usepackage{nccmath}
\usepackage{empheq} 
\usepackage[normalem]{ulem} 
\usepackage{float}
\usepackage{xcolor}

\definecolor{blue-violet}{rgb}{0.54, 0.17, 0.89} 

\let\oldtimes\times  
\renewcommand\times{{\oldtimes}}

\usepackage[cmyk,dvipsnames]{xcolor}
\definecolor{darkorchid}{HTML}{bf3eff}

\begin{document}
    
\author{Zirui Zhang}

\affiliation{Department of Physics, \href{https://ror.org/03893we55}{Zhejiang Sci-Tech University}, Hangzhou 310018, China}

\author{Shiwei Zhu}
\author{Ruixiao Ma}
\author{Qiuyao Zhang}
\author{Xiaoping Wu}

\affiliation{Department of Physics, \href{https://ror.org/03893we55}{Zhejiang Sci-Tech University}, Hangzhou 310018, China}

\author{Changsheng Song}
\email[Contact author:]{cssong@zstu.edu.cn}

\affiliation{Department of Physics, \href{https://ror.org/03893we55}{Zhejiang Sci-Tech University}, Hangzhou 310018, China}

\affiliation{Zhejiang Key Laboratory of Quantum State Control and Optical Field Manipulation, \href{https://ror.org/03893we55}{Zhejiang Sci-Tech University}, Hangzhou 310018, China}

\title
{Topological spin textures in 2D altermagnetic chromium chalcogenides: Interplay between magnetic frustration and Dzyaloshinskii-Moriya interaction}
	
\date{\today}
\begin{abstract}

Altermagnetism has recently emerged as a distinct magnetic paradigm, yet the exploration of topological spin textures within two-dimensional (2D) altermagnets remains limited. 
Using first-principles calculations and atomistic spin simulations, we investigate monolayer $\mathrm{Cr_2X_2}$ and Janus monolayer $\mathrm{Cr_2XY}$ altermagnets ($\mathrm{X, Y}=\mathrm{O, S, Se, Te}$, and $\mathrm{X \neq Y}$). 
Both systems exhibit strong intrinsic exchange frustration arising from competition between nearest-neighbor and further-neighbor exchange interactions. In Janus $\mathrm{Cr_2XY}$, breaking the out-of-plane mirror symmetry additionally activates a substantial in-plane Dzyaloshinskii-Moriya interaction (DMI), mediated by the strong spin-orbit coupling of heavy chalcogen ligands.
The interplay between exchange frustration and DMI gives rise to distinct frustration- and DMI-dominated regimes, while magnetic anisotropy determines the preferred spin orientation, leading to a diverse range of non-collinear states. 
Systematic phase-space scans and geodesic nudged elastic band (GNEB) calculations reveal two distinct mechanisms governing this interplay: a synergistic effect, in which frustration lowers the critical DMI for the onset of non-collinear states, and a competitive effect, in which frustration favors high-topological-charge states while DMI favors low-charge states. 
Our results establish a general framework for understanding the interplay between frustration and DMI and guiding the design of topological spin textures in two-dimensional altermagnets. 

\end{abstract}

\maketitle

\section{INTRODUCTION}

Topological spin textures, such as skyrmions and bimerons, have attracted intense interest owing to their non-trivial topology, rich internal degrees of freedom, and potential applications in ultra-dense, low-power spintronic devices~\cite{fert2013skyrmions,fert2017magnetic,bimeron_PRB2019,back20202020,gobel2021beyond,he2023all,Rybakov2025}. Their topological character provides robustness against continuous deformations and enables a variety of topological states, including high-topological-charge (high-$Q$) configurations~\cite{hassan2024dipolar, niu2025magnetic, shiwei2026}, making them versatile platforms for exploring emergent topological phenomena and developing next-generation spintronic technologies. Their formation and stability critically depend on competing magnetic interactions, most notably DMI arising from spin–orbit coupling (SOC) in non-centrosymmetric environments~\cite{DZYALOSHINSKY1958241, PhysRev.120.91} and magnetic frustration from competing exchange interactions~\cite{PhysRevB.101.045416, paul2020role, von2017enhanced}, together with other competing mechanisms such as dipolar interactions and higher-order exchange interactions~\cite{Ezawa2010giant,paul2020role}. 
Despite extensive studies of these mechanisms in conventional ferromagnets~\cite{roessler2006spontaneous,leonov2016properties,wijethunga2025phase,wang2018theory,wu2021size,hu2022theory,wu2022nematic,Okubo2012,leonov2015multiply,von2017enhanced} and evidence for their analogous roles in antiferromagnets \cite{zhang2016antiferromagnetic,legrand2020room,dou2023theoretical}, 
previous work has largely treated the two mechanisms separately, and
how their interplay governs the formation and phase behavior of topological magnetic states in emerging magnetic phases remains poorly understood.

Altermagnetism (AM) has recently emerged as a distinct magnetic paradigm, characterized by zero net magnetization and momentum-dependent spin splitting arising from alternating spin polarization in real and reciprocal spaces~\cite{vsmejkal2022emerging,vsmejkal2022beyond,zhu2024observation,krempasky2024altermagnetic}. Extensive studies have focused on the electronic structure, anomalous Hall effect, and spin transport of altermagnets~\cite{lee2024broken,gonzalez2023spontaneous,wu2024valley}, but topological spin textures within AM backgrounds remain largely unexplored. 
To address this question, 2D chromium chalcogenides $\mathrm{Cr_2X_2}$ ($\mathrm{X} = \mathrm{O, S, Se}$, and $\mathrm{Te}$) serve as a compelling prototype. Previous research has established the AM ground state and electronic properties of these monolayers, including the Janus member $\mathrm{Cr_2SO}$~\cite{guo2023quantum,guo2023piezoelectric}.
Chiral spin textures have been predicted in related Janus chromium compounds, but only in their ferromagnetic (FM) counterparts~\cite{shen2022strain, cui2020strain}.
Notably, these candidate altermagnets can be described as square-lattice Heisenberg antiferromagnets with exclusively antiferromagnetic (AFM), including the further-neighbor ones, so that they realize a frustrated square-lattice $J_1–J_2–J_3$ model~\cite{weihong1991square,qiao2026phase} in which strong frustration arises from same-sign competition rather than from opposing FM and AFM couplings.
However, the out-of-plane mirror symmetry ($\mathcal{M}_{z}$) of $\mathrm{Cr_2X_2}$ forbids any net in-plane DMI ($D_{\parallel}$), so that exchange frustration is left as the only remaining in-plane driving force toward non-collinear order, limiting the possibility of stabilizing chiral spin textures.
The asymmetric Janus configuration $\mathrm{Cr_2XY}$ ($\mathrm{X, Y}=\mathrm{O, S, Se, Te}$, and $\mathrm{X \neq Y}$) naturally breaks this mirror symmetry,
thereby activating a sizable $D_{\parallel}$ while retaining the intrinsic exchange frustration inherited from the parent material. 
Taken together, $\mathrm{Cr_2XY}$ monolayers provide a platform in which strong exchange frustration and a tunable $D_{\parallel}$ coexist, allowing us to ask whether and how their interplay stabilizes topological skyrmions and bimerons at zero magnetic field within an AM background.

\begin{figure*}[ht]
	\centering
	\includegraphics[width=0.9\linewidth]{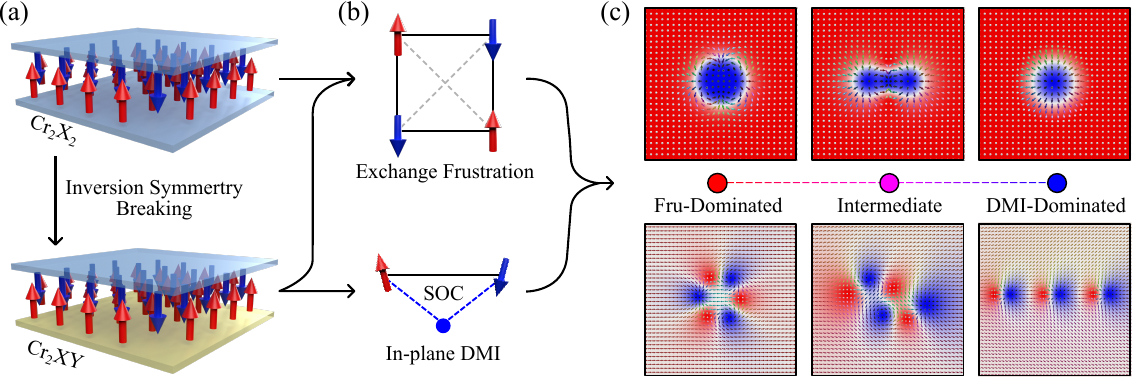}
	\caption{\label{fig_1}
Schematic illustration of the formation mechanisms and classification of topological spin textures in monolayers $\mathrm{Cr_2X_2}$ and Janus $\mathrm{Cr_2XY}$ altermagnets.
(a) Spin configurations of $\mathrm{Cr_2X_2}$ (symmetric, with out-of-plane mirror symmetry $\mathcal{M}_{z}$) and Janus $\mathrm{Cr_2XY}$ (asymmetric, $\mathcal{M}_{z}$ broken). Different colors represent distinct ligand atomic layers, red and blue arrows denote the spin directions. 
(b) Two coexisting driving forces: exchange frustration arising from competing neighbor exchange interactions, and in-plane DMI ($D_\parallel$) activated by broken $\mathcal{M}_{z}$ symmetry combined with strong spin-orbit coupling (SOC).
(c)  Classification of the resulting topological spin textures into three regimes based on the relative strength of frustration and DMI: frustration-dominated, intermediate, and DMI-dominated.
            }
\end{figure*}

In this work, by combining first-principles calculations and atomistic spin-model simulations, we systematically investigate the lattice stability, spin-Hamiltonian parameters, and topological spin textures in monolayer $\mathrm{Cr_2X_2}$ and Janus monolayer $\mathrm{Cr_2XY}$ altermagnets. The broken $\mathcal{M}_{z}$ in Janus $\mathrm{Cr_2XY}$ activates a substantial in-plane DMI, $D_{\parallel}$ (see Fig.~\ref{fig_1}(a)), while competing multi-neighbor exchange interactions give rise to strong exchange frustration. Atomistic spin-dynamics simulations reveal that, under zero external field, a variety of topological spin textures, including bimeron clusters, bimeron chains, and isolated skyrmions, can be stabilized within the AM background. Crucially, by constructing parameter-space phase diagrams as functions of frustration parameters and $D_{\parallel}$, we unveil two coexisting mechanisms: a synergistic effect, in which frustration softens the collinear AM order and lowers the $D_{\parallel}$ threshold for forming topological spin configurations, and an opposing effect, in which frustration favors high-$Q$ states while $D_{\parallel}$ favors single-$Q$ skyrmions. These findings establish the frustration–DMI balance as a handle for tuning the topological charge and morphology of spin textures in 2D altermagnets, and identify $\mathrm{Cr_2X_2}$ as a platform for field-free topological spin textures.

\section{METHODS}
Our first-principles calculations are performed within density functional theory (DFT) as implemented in the Vienna \textit{Ab initio} Simulation Package (VASP)~\cite{hafner2008ab}. The projector augmented wave (PAW) method is used to describe the interaction between ionic cores and valence electrons, and the exchange–correlation potential is treated using the Perdew–Burke–Ernzerhof (PBE) functional within the generalized gradient approximation (GGA). A vacuum layer more than 15~\AA{} is applied along the out-of-plane direction to avoid interactions between periodic images. The GGA+$U$ method with $U = 3.5$ eV is used for the Cr $3d$ electrons from the structural optimization stage onward~\cite{guo2023quantum,guo2023piezoelectric}. The plane-wave cutoff energy is set to 500 eV. The Brillouin zone is sampled using a $9 \times 9 \times 1$ Monkhorst–Pack $k$-point mesh. The convergence criteria for total energy and atomic forces are $10^{-8}$ eV and $0.001$ eV/\AA, respectively. Phonon spectra are calculated using the PHONOPY package with a $2 \times 2 \times 1$ supercell~\cite{phonopy-phono3py-JPCM} based on the finite-displacement method. To evaluate the Néel temperature ($T_{\mathrm{N}}$), classical Monte Carlo simulations are carried out using the SpinMC.jl package~\cite{buessen2025spinmc} based on the derived Heisenberg spin model, where a $16 \times 16 \times 1$ periodic supercell is adopted. 
$T_{\mathrm{N}}$ is identified from the peak position of the specific heat $C_{\mathrm{V}}$.

Atomistic spin-dynamics simulations were performed by solving the Landau–Lifshitz–Gilbert (LLG) equation using the SPIRIT code~\cite{PhysRevB.99.224414} to obtain the equilibrium spin configurations under zero external magnetic field at zero temperature, with a Gilbert damping parameter of $\alpha = 0.3$.
Furthermore, the GNEB method was employed to determine the minimum-energy paths for the collapse of topological spin textures into the uniform AM ground state and the corresponding energy barriers~\cite{bessarab2015method}.
The atomistic spin simulations, including both the LLG relaxation and the GNEB calculations, are performed on a $100 \times 100 \times 1$ periodic supercell, and the total number of iteration steps set to $5 \times 10^5$ to ensure that the system fully relaxes into stable configurations.

\begin{figure*}[ht]
	\centering
	\includegraphics[width=1.0\linewidth]{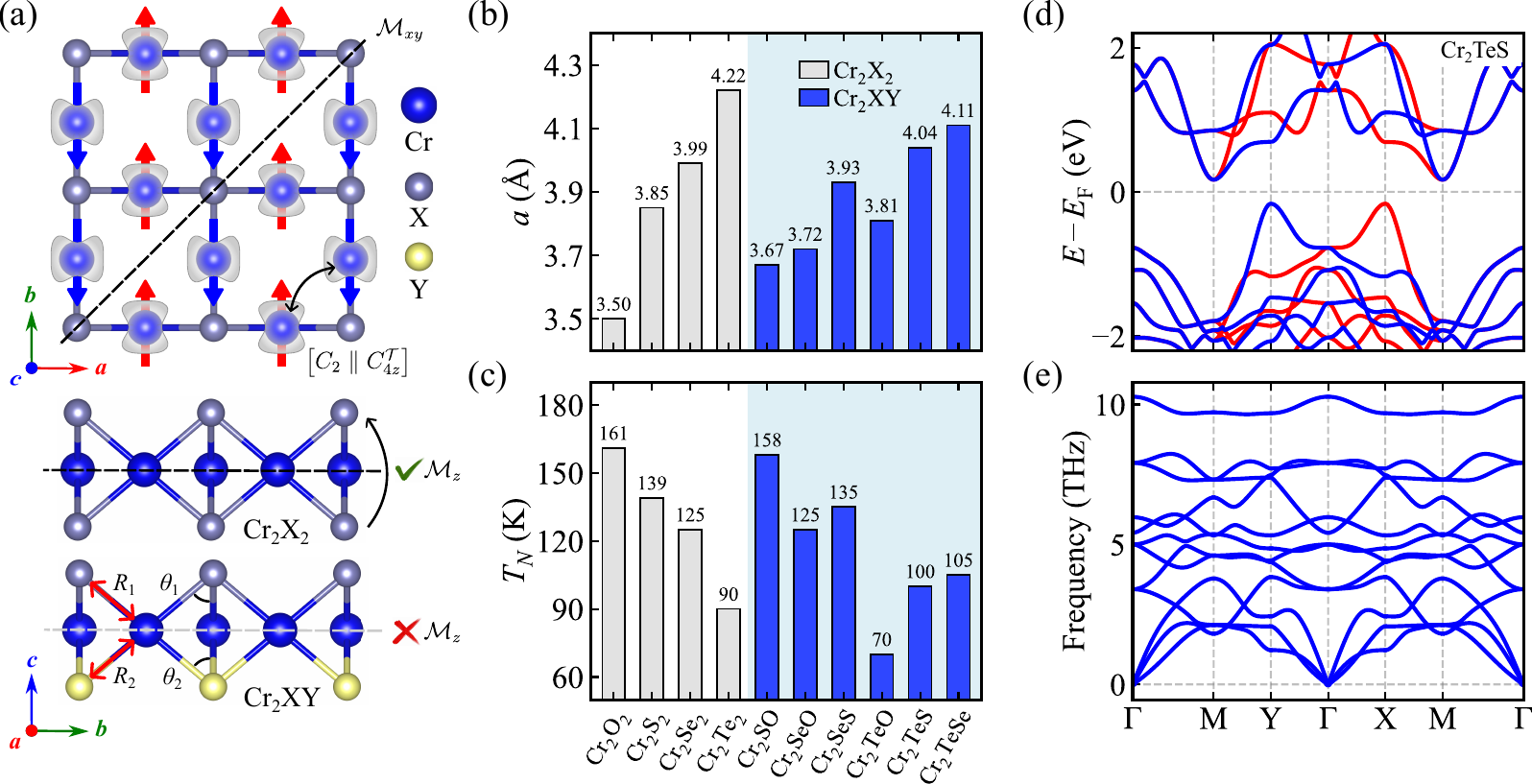}
	\caption{\label{fig_2}
Structural, magnetic, and electronic properties of the 2D AM $\mathrm{Cr_2X_2}$ and Janus Cr$_2$XY monolayers.
(a) Crystal structure, spin-density distribution, and schematic symmetry operations connecting the two magnetic sublattices in $\mathrm{Cr_2X_2}$/Cr$_2$XY monolayers.
(b) Calculated lattice constants $a$ and
(c) predicted N\'eel temperatures $T_{\mathrm N}$ of the $\mathrm{Cr_2X_2}$ and Cr$_2$XY families, where gray and blue bars denote $\mathrm{Cr_2X_2}$ and Cr$_2$XY materials, respectively.
(d) Spin-resolved band structure of the representative Janus altermagnet Cr$_2$TeS, where the red and blue curves denote opposite spin channels.
(e) Phonon spectrum of Cr$_2$TeS, confirming its dynamical stability.
    }
\end{figure*}

\section{Results and Discussion}

The formation of topological spin textures and its microscopic origins in monolayer $\mathrm{Cr_2X_2}$ and Janus $\mathrm{Cr_2XY}$ are schematically illustrated in Fig.~\ref{fig_1}. In symmetric $\mathrm{Cr_2X_2}$ monolayers, the out-of-plane mirror symmetry $\mathcal{M}_z$ strictly forbids any net $D_{\parallel}$. The asymmetric Janus configuration of $\mathrm{Cr_2XY}$ breaks this symmetry [Fig.~\ref{fig_1}(a)] and, together with strong SOC, activates a substantial $D_{\parallel}$. 
The interplay between $D_{\parallel}$ and intrinsic exchange frustration [Fig.~\ref{fig_1}(b)] stabilizes various non-collinear spin configurations, which can be classified into three regimes in the frustration–DMI parameter space: frustration-dominated, intermediate, and DMI-dominated [Fig.~\ref{fig_1}(c)].
To quantitatively map these mechanisms and identify stable topological states, a comprehensive understanding of the lattice structures, electronic properties, and magnetic interactions is required. In the following, we comprehensively explore these candidate materials, beginning with their structural configurations and dynamical stability.

\subsection*{A. Structural, magnetic, and electronic properties \label{subsec:IIIA}}

We first examine the crystal symmetry and local coordination environment of 2D $\mathrm{Cr_2X_2}$ and Janus $\mathrm{Cr_2XY}$ monolayers. As illustrated in Fig.~\ref{fig_2}(a), both families consist of a central $\mathrm{Cr}$ layer sandwiched between top and bottom $\mathrm{X/Y}$ ligand layers. 
Both families retain the fourfold rotational symmetry $C_4$ and the diagonal vertical mirror planes, whereas the horizontal mirror plane $\mathcal{M}_z$ exists only in the symmetric $\mathrm{Cr_2X_2}$ monolayers and is broken in the Janus $\mathrm{Cr_2XY}$ monolayers.
In the AM ground state, the opposite spins partition the lattice into two distinct magnetic sublattices, which are connected by the combined $\left[C_{2} \parallel C_{4z}^{\mathcal{T}}\right]$ symmetry operation, ensuring macroscopically compensated magnetization. 
Each Cr atom is coordinated by four neighboring ligands, forming an edge-sharing $\mathrm{CrX_4}$ or $\mathrm{CrX_2Y_2}$ tetrahedron. For symmetric $\mathrm{Cr_2X_2}$ monolayers (space group $P4/mmm$), the out-of-plane mirror symmetry $\mathcal{M}_z$ requires the upper and lower ligand shells to be identical, resulting in equal bond lengths ($R_1 = R_2$) and bond angles ($\theta_1 = \theta_2$), consistent with the local local coordination symmetry of $D_{2d}$ in the Cr ions. In contrast, replacing the top ligand layer lowers the symmetry to the polar $P4mm$ space group in Janus $\mathrm{Cr_2XY}$ monolayers, breaking $\mathcal{M}_z$ and reducing the local local coordination symmetry of Cr to $C_{2v}$. This symmetry breaking provides the structural prerequisite for activating the in-plane DMI discussed in Sec.~\ref{subsec:IIIB} B. 
It also induces a vertical electric dipole moment and structural distortions (see Sec.~S1 in the Supplemental Material (SM)~\cite{supplmat}). For instance, materials with a large atomic-radius mismatch, such as Janus $\mathrm{Cr_2TeO}$, exhibit a pronounced asymmetry between $R_1$ (2.773~\AA) and $R_2$ (2.156~\AA), reflecting the structural relaxation required to accommodate the different ligand sizes.

As shown in Fig.~\ref{fig_2}(b), the optimized lattice constants increase with the atomic size of the ligand elements, ranging from 3.50~\AA{} for $\mathrm{Cr_2O_2}$ to 4.22~\AA{} for $\mathrm{Cr_2Te_2}$, in good agreement with previous reports for $\mathrm{Cr_2X_2}$ and $\mathrm{Cr_2XY}$ monolayers~\cite{guo2023quantum,guo2023piezoelectric}.
For the Janus materials, the lattice constants consistently fall between those of their corresponding symmetric parent materials, reflecting a regular structural evolution across the chemical series. 
Using the exchange parameters derived from DFT, the calculated $T_{\mathrm{N}}$ are shown in Fig.~\ref{fig_2}(c). They range from 70~K for Janus $\mathrm{Cr_2TeO}$ to 161~K for symmetric $\mathrm{Cr_2O_2}$. 
The highest $T_{\mathrm{N}}$ in $\mathrm{Cr_2O_2}$ is consistent with its shortest Cr–Cr bonds and the largest $|J_1|$, whereas the anomalously low value in $\mathrm{Cr_2TeO}$ reflects its extreme structural distortion, which correlates with $|J_2'|$ and enhances frustration (see Tables S I, S II and detailed temperature-dependent heat-capacity curves in Sec.~S2 of the SM~\cite{supplmat}).
All curves exhibit prominent $\lambda$-type peaks, signifying the second-order magnetic phase transition from the AM phase to the paramagnetic (PM) state.

We next analyze the electronic structure and dynamical stability of these monolayers. Despite their zero net magnetization, the breaking of space-time inversion symmetry ($\mathcal{PT}$) in the AM state lifts the Kramers spin degeneracy~\cite{krempasky2024altermagnetic}, producing momentum-dependent spin splitting along specific paths in the Brillouin zone, as demonstrated by the spin-resolved band structure of the representative Janus material $\mathrm{Cr_2TeS}$ [Fig.~\ref{fig_2}(d)]. 
The same altermagnetic splitting is found for all other monolayers (see Sec. S3 of the SM~\cite{supplmat}).
The phonon spectra confirm the dynamical stability of all materials except Janus $\mathrm{Cr_2TeO}$ (Fig.~\ref{fig_2}(e) and Sec.~S3 of the SM~\cite{supplmat}).
Although the pronounced structural mismatch in Janus $\mathrm{Cr_2TeO}$ induces soft phonon modes, indicating dynamical instability associated with internal structural stress, we retain it as an illustrative limiting case to establish complete and continuous physical trends across this 2D AM family.
With the structural and dynamical properties established, we next extract the magnetic interaction parameters that govern their spin behavior.

\subsection*{B. Spin-Hamiltonian parameters \label{subsec:IIIB}}

To describe the magnetic interactions, we adopt the following spin Hamiltonian:
\begin{equation}\label{eq:hamiltonian}
\begin{aligned}
\mathcal{H} = & -\sum_{\langle i,j \rangle} J_1 (\mathbf{S}_i \cdot \mathbf{S}_j) - \sum_{\langle\langle i,j \rangle\rangle} J_2 (\mathbf{S}_i \cdot \mathbf{S}_j) \\
    & - \sum_{\langle\langle i,j \rangle\rangle'} J_2' (\mathbf{S}_i \cdot \mathbf{S}_j) - \sum_{\langle\langle\langle i,j \rangle\rangle\rangle} J_3 (\mathbf{S}_i \cdot \mathbf{S}_j) \\
    & - \sum_{\langle i,j \rangle} \mathbf{D}_{ij} \cdot (\mathbf{S}_i \times \mathbf{S}_j) - \sum_{i} K (S_i^z)^2.
\end{aligned}
\end{equation}

\begin{figure}[htb]
	\centering
	\includegraphics[width=0.9\linewidth]{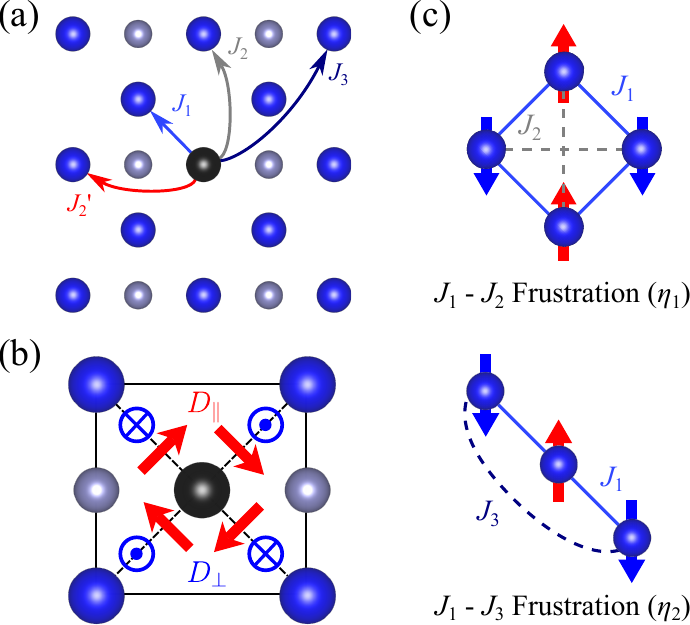}
	\caption{\label{fig_3}
Magnetic interactions and exchange frustration in $\mathrm{Cr_2X_2}$ and Janus $\mathrm{Cr_2XY}$ monolayers.
(a) Schematic of Heisenberg exchange interactions $J_1$ (nearest-neighbor), $J_2$ and $J_2'$ (second-nearest-neighbor, distinguished by the presence or absence of ligand mediation), and $J_3$ (third-nearest-neighbor).
(b) Nearest-neighbor DMI vector, showing the in-plane ($D_{\parallel}$) and out-of-plane ($D_{\perp}$) components.
(c) Exchange frustration arising from competing AFM exchange interactions: $J_1$--$J_2$ ($\eta_1$) and $J_1$--$J_3$ ($\eta_2$) cases. Solid and dashed lines represent satisfied and unsatisfied exchange bonds, respectively.
    }
\end{figure}

Here, $J>0$ denotes FM Heisenberg exchange, while $J<0$ corresponds to AFM exchange. The parameters $J_1$, $J_2$, $J_2'$, and $J_3$ represent the nearest-neighbor (NN), second-NN, ligand-mediated second-NN, and third-NN exchange interactions, respectively, as illustrated in Fig.~\ref{fig_3}(a). 
$\mathbf{S}_i$ and $\mathbf{S}_j$ are the spin vectors at $i$ and $j$ sites. 
$\mathbf{D}_{ij}$ is the DMI vector, with only the NN contribution retained. Following the symmetry constraints and Moriya's rules~\cite{PhysRev.120.91}, it can be written as $\mathbf{D}_{ij} = D_{\parallel} (\hat{u}_{ij} \times \hat{z}) + D_{\perp} \hat{z}$, where $D_{\parallel}$ and $D_{\perp}$ denote the in-plane and out-of-plane DMI components, respectively, $\hat{u}_{ij}$ is the in-plane unit vector pointing from site $i$ to $j$, and $\hat{z}$ is the unit vector normal to the monolayer plane. Under our convention, $D_{\parallel}>0$ favors the clockwise (CW) spin chirality, while $D_{\perp}>0$ corresponds to a DMI vector pointing along the $+z$ direction (see Sec.~S4 in the SM~\cite{supplmat}). Their spatial distributions are schematically illustrated in Fig.~\ref{fig_3}(b). 
$K$ denotes the single-ion anisotropy, with $K>0$ and $K<0$ corresponding to out-of-plane (OOP) and in-plane (IP) anisotropy, respectively. All spin-Hamiltonian parameters are extracted from DFT total-energy differences between collinear and non-collinear spin configurations (see Sec.~S4 of the SM for computational details~\cite{supplmat}).

\begin{figure}[tb]
	\centering
	\includegraphics[width=1\linewidth]{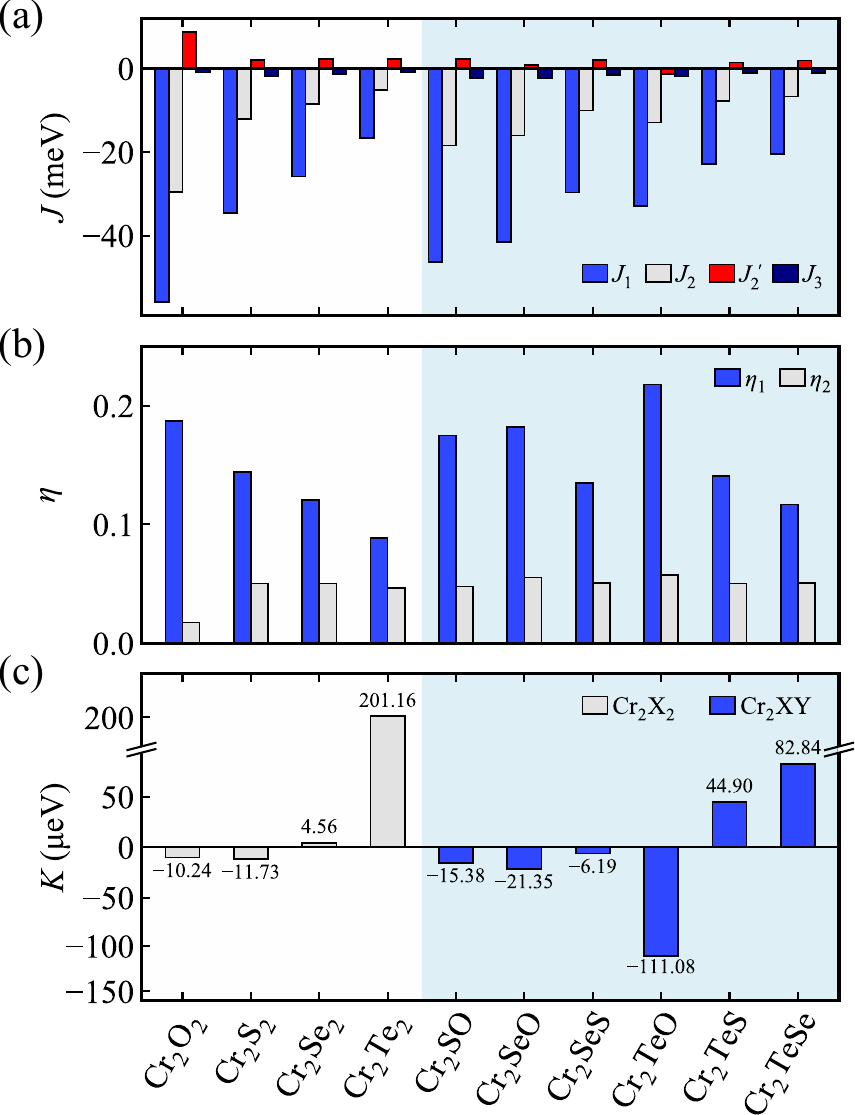}
	\caption{\label{fig_4}
Exchange interactions, frustration parameters, and single-ion anisotropies of the $\mathrm{Cr_2X_2}$ and Janus $\mathrm{Cr_2XY}$ monolayers.
(a) Calculated Heisenberg exchange coefficients $J_1$, $J_2$, $J_2'$, and $J_3$.
(b) Frustration parameters.
(c) Single-ion anisotropy coefficients $K$.
    }
\end{figure}

As shown in Fig.~\ref{fig_4}(a), $J_1$, $J_2$, and $J_3$ in the $\mathrm{Cr_2X_2}$ and $\mathrm{Cr_2XY}$ systems are predominantly AFM. The dominant $J_1<0$ favors antiparallel alignment between NN spins, leading to parallel alignment within each magnetic sublattice. In contrast, the further-neighbor interactions $J_2,J_3<0$ favor antiparallel alignment within the same sublattice, competing with the $J_1$-driven ordering and giving rise to intrinsic exchange frustration. 
The ligand-mediated second-NN interaction $J_2'$ is positive in most compounds (see Table S~II), partially opposing this frustration, but is much weaker than $J_2$. Because each Cr ion couples to two inequivalent second neighbors of each type, we define the effective second-NN exchange as $J_2^{\mathrm{eff}}=(J_2+J_2')/2$, as schematically illustrated in Fig.~\ref{fig_3}(c). The resulting $J_2^{\mathrm{eff}}$ remains negative for all systems, preserving the AFM character of the effective second-NN interaction and the associated frustration. We therefore introduce two frustration ratios,
\begin{equation}
\eta_1 = \frac{J_2^{\text{eff}}}{J_1}, \quad \eta_2 = \frac{J_3}{J_1},
\end{equation}
where $\eta_1$ and $\eta_2$ quantify the relative strengths of the $J_1$--$J_2$ and $J_1$--$J_3$ frustration, respectively. Both ratios are positive because $J_1$, $J_2^{\rm eff}$, and $J_3$ are all AFM.

\begin{figure}[tb]
	\centering
	\includegraphics[width=1\linewidth]{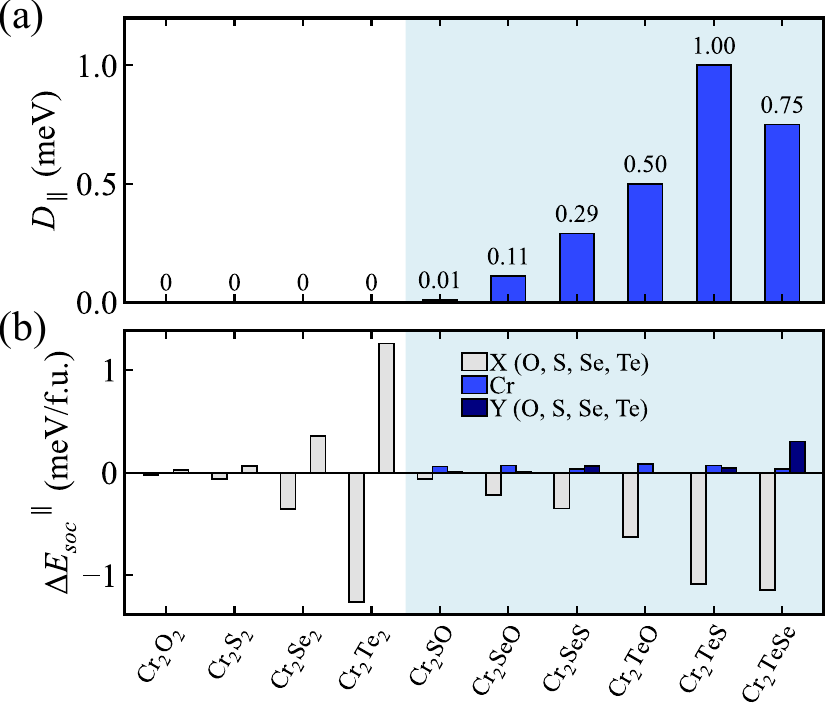}
	\caption{\label{fig_5}
In-plane DMI and its atomically origin in $\mathrm{Cr_2X_2}$ and Janus $\mathrm{Cr_2XY}$ monolayers. 
(a) Calculated in-plane DMI component $D_\parallel$.
(b) Atomically resolved spin-orbit coupling (SOC) energy difference $\Delta E_{\mathrm{SOC}}^{\parallel}$ decomposed into contributions from $\mathrm{Cr}$ and chalcogen ($\mathrm{X, Y}$) layers.
    }
\end{figure}

The calculated frustration parameters $\eta_1$ and $\eta_2$ are plotted in Fig.~\ref{fig_4}(b). 
Across all studied materials, $\eta_1$ ranges from 0.09 to 0.22 and consistently dominates over $\eta_2$, which remains relatively small (0.02–0.06).
Although $\eta_2$ is smaller than $\eta_1$, $J_3$ carries approximately twice the weight of $J_2$ in the exchange stiffness of a square lattice~\cite{rybakov2022magnetic, sallermann2023stability, zhu2026beatingmicromagneticlimitsskyrmion}, making its effective contribution non-negligible. 
Thus, both $\eta_1$ and $\eta_2$ contribute to the overall exchange frustration at the sublattice level, enabling non-collinear spin states.

Next, the single-ion anisotropy $K$ plotted in Fig.~\ref{fig_4}(c) plays a vital role in determining the preferred spin orientation. A clear ligand-dependent evolution of the magnetic anisotropy is observed. In the centrosymmetric $\mathrm{Cr_2X_2}$ family, the system undergoes a transition from IP to OOP anisotropy as the ligand changes from the lighter elements (O, S) to the heavier ones (Se, Te). This evolution is driven by the increasing SOC strength of the heavier ligands, which competes with crystal-field splitting to reorient the magnetization~\cite{wang1993first,hu2013control}. 
A similar overall trend emerges in the Janus $\mathrm{Cr_2XY}$ monolayers, where combinations involving lighter ligands predominantly exhibit IP anisotropy, while heavier-ligand combinations tend to favor an OOP, as exemplified by $\mathrm{Cr_2TeS}$ and $\mathrm{Cr_2TeSe}$. Nevertheless, the anisotropy does not evolve monotonically with ligand mass, reflecting the additional influence of the local coordination environment. In particular, $\mathrm{Cr_2TeO}$ exhibits an exceptionally large IP anisotropy of $K=-111.08~\mu\mathrm{eV/Cr}$, which can be attributed to the pronounced structural distortion of its coordination tetrahedra.

\begin{figure*}[tb]
	\centering
	\includegraphics[width=1\linewidth]{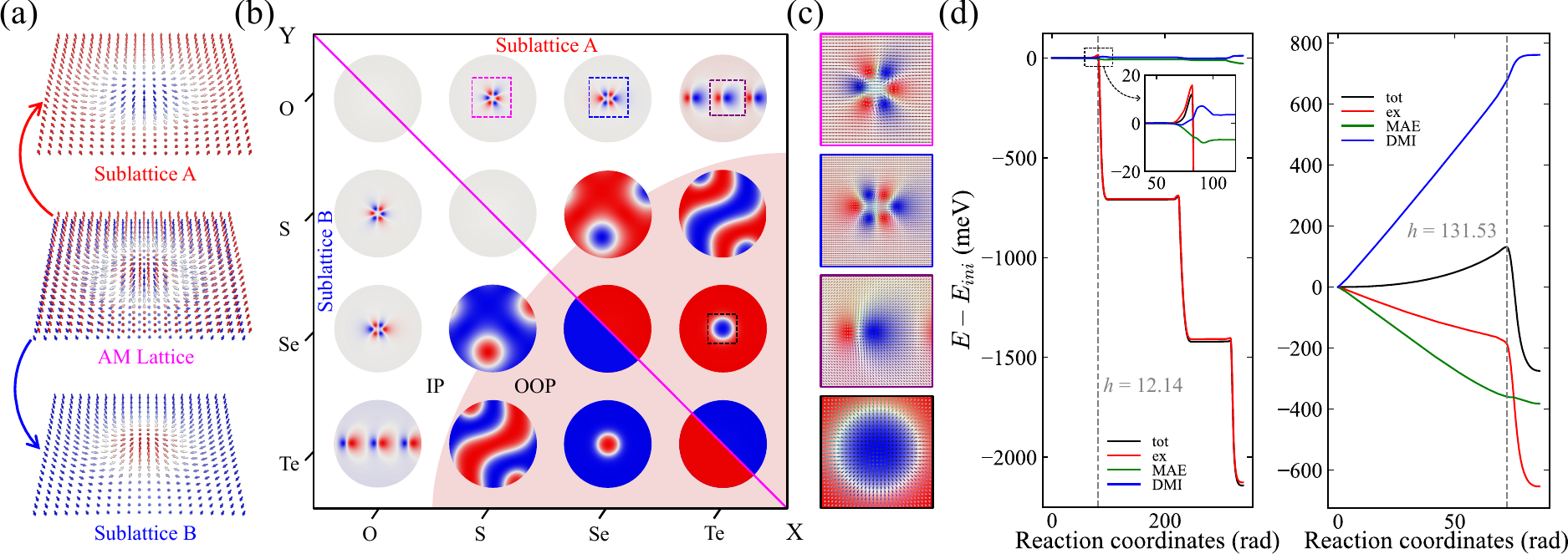}
	\caption{\label{fig_6}
Topological spin textures and minimum energy paths of $\mathrm{Cr_2X_2}$ and Janus Cr$_2$XY monolayers.
(a) Decomposition of the AM lattice into two single-sublattice networks (Sublattice A and Sublattice B) . The subsequent spin-dynamics simulations and analyses are performed based on individual sublattices.
(b) Zero-field magnetic phase diagram. Pink and transparent regions indicate materials with OOP and IP magnetic anisotropy, respectively. The solid magenta diagonal line divides the phase space into Sublattice A and Sublattice B.
(c) Enlarged spin configurations for representative materials: $\mathrm{Cr_2SO}$ (bimeron cluster), $\mathrm{Cr_2SeO}$ (bimeron cluster), $\mathrm{Cr_2TeO}$ (bimeron chain), and $\mathrm{Cr_2TeSe}$ (skyrmion).
(d) Minimum energy paths for topological collapse in $\mathrm{Cr_2SeO}$ (left) and $\mathrm{Cr_2TeSe}$ (right). The total energy (tot) and individual contributions from exchange (ex), magnetic anisotropy energy (MAE), and Dzyaloshinskii-Moriya interaction (DMI) are shown. The dashed vertical lines indicate the calculated energy barriers $h$ (in meV); Left panel includes a magnified inset of the saddle point (black dashed box).
    }
\end{figure*}

Turning to the DMI, although both $D_{\parallel}$ and $D_{\perp}$ are included in our calculations, we focus on $D_{\parallel}$ below.
The out-of-plane component takes values between $0.27$ and $1.57$ meV, which is comparable to or even larger than $D_{\parallel}$, but it is $D_{\parallel}$ that fixes the rotational sense of the in-plane and Néel-type textures realized in this work.
Consistent with the symmetry analysis in Sec.~\ref{subsec:IIIA} A, $D_{\parallel}$ vanishes in the symmetric $\mathrm{Cr_2X_2}$ monolayers and becomes finite in the Janus $\mathrm{Cr_2XY}$ configuration, as summarized in Fig. ~\ref{fig_5}(a) and Table S II of the SM~\cite{supplmat}. 
The magnitude of $D_{\parallel}$ generally increases with the atomic number of the heavier chalcogen, from $0.01$~meV in $\mathrm{Cr_2SO}$ to a maximum of $1.00$~meV in $\mathrm{Cr_2TeS}$, before slightly decreasing to $0.75$~meV in $\mathrm{Cr_2TeSe}$. 
This trend, together with the atom-resolved SOC energy differences ($\Delta E_{\mathrm{SOC}}$) in Fig.~\ref{fig_5}(b), indicates that the dominant contribution comes from the heavy chalcogen atoms (Se and Te) rather than Cr. 
This behavior is consistent with the Fert-Levy mechanism~\cite{fert1980role}, where strong ligand SOC mediates asymmetric exchange between the Cr $3d$ spins and generates a finite $D_{\parallel}$.

\subsection*{C. Zero-field topological spin textures \label{subsec:IIIC}}

Having determined the parameters of exchange, anisotropy, and DMI, we next examine the spin textures stailzed by these interactions.
To systematically investigate the formation and diversity of topological spin textures in our AM monolayers, we perform atomistic spin simulations under zero external field based on Eq.~\ref{eq:hamiltonian}. The AM lattice consists of two interpenetrating sublattices with antiparallel spin orientations, resulting in zero net magnetization. Directly evaluating the topological charge $Q$ on the full bipartite lattice would cause the opposite contributions from the two sublattices to cancel. Therefore, as illustrated in Fig.~\ref{fig_6}(a), we decompose the lattice into two independent single-sublattice magnetic networks, denoted as sublattice A and sublattice B, and perform all subsequent topological analyses at the single-sublattice level.

\begin{figure*}[tb]
	\centering
	\includegraphics[width=1\linewidth]{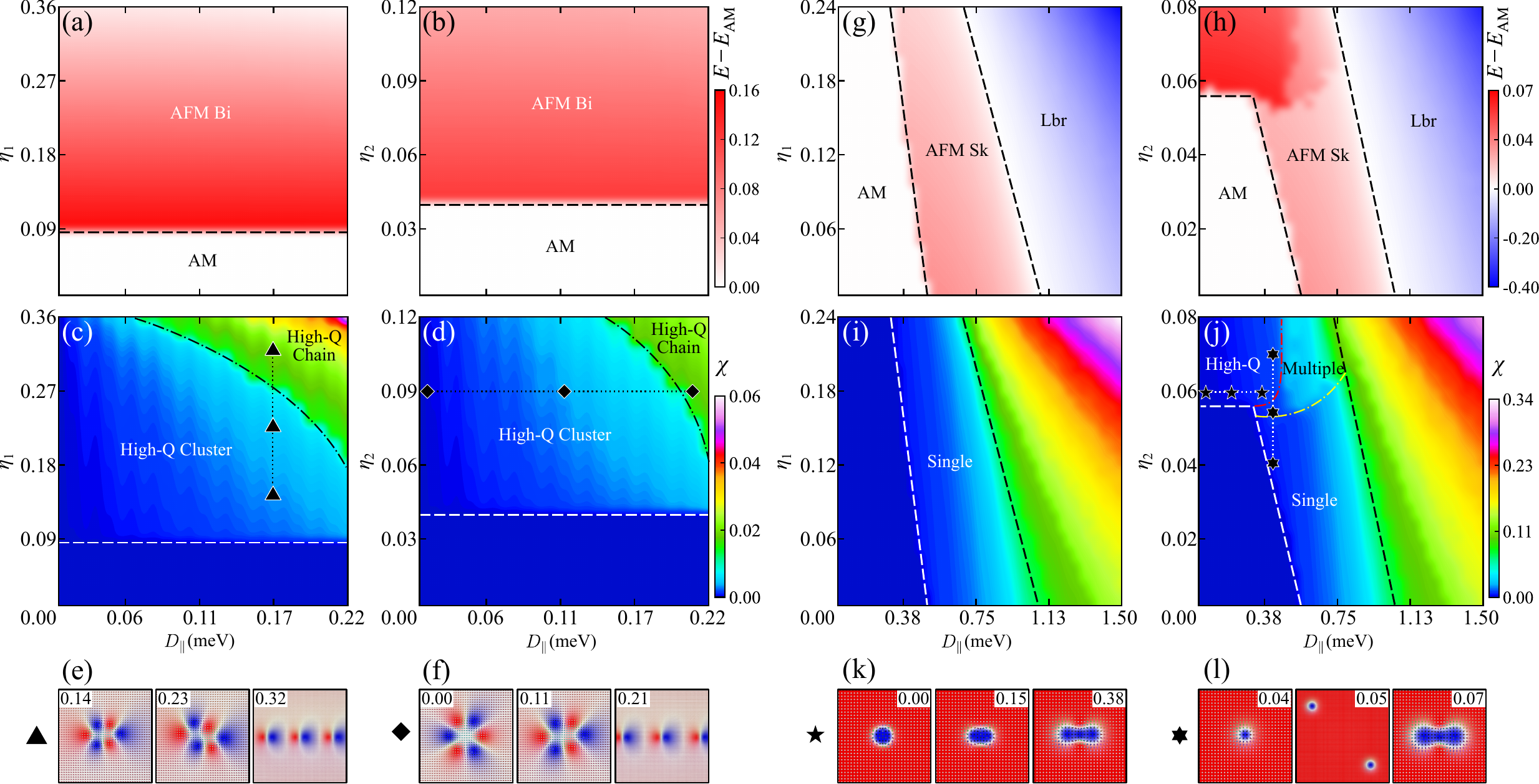}
	\caption{\label{fig_7}
Phase diagrams and topological spin texture evolution driven by exchange frustration and in-plane DMI. 
(a--f) Frustration-dominated material $\mathrm{Cr}_2\mathrm{SeO}$. (g--l) DMI-dominated material $\mathrm{Cr}_2\mathrm{TeSe}$. 
(a, b, g, h) Energy difference $E - E_{\mathrm{AM}}$ (in meV) in the $\eta_1-D_{\parallel}$ and $\eta_2-D_{\parallel}$ space. Dashed lines mark the phase boundaries between AM states, topological phase regions (bimeron / skyrmion), and labyrinthine (Lbr) states. 
(c, d, i, j) Phase diagrams of the chiral correlation $\chi$. Dash-dotted lines distinguish sub-phases with different topological configurations. Dotted lines indicate selected parameter trajectories. 
(e, f, k, l) Sublattice spin textures extracted along the trajectories in (c, d, i, j), with sampling points denotes by symbols.
    }
\end{figure*}

By relaxing the LLG dynamics to equilibrium, as shown in Fig.~\ref{fig_6}(b), we obtain the zero-field magnetic phase diagram for all $\mathrm{Cr_2X_2}$ and Janus $\mathrm{Cr_2XY}$ monolayers. 
Owing to their distinct intrinsic elemental component, the Janus monolayers realize different textures within the phase diagram.
The phase diagram is constructed at the single-sublattice level, with the upper-right and lower-left regions corresponding to Sublattices A and B, respectively. The representative states reconstructed on the full bipartite lattice in Fig.~\ref{fig_6}(c) are selected from the textures of $\mathrm{Cr_2SO}$, $\mathrm{Cr_2SeO}$, $\mathrm{Cr_2TeO}$ and $\mathrm{Cr_2TeSe}$ in the corresponding panels of Fig.~\ref{fig_6}(b). 
For the four centrosymmetric $\mathrm{Cr_2X_2}$ monolayers, $\mathcal{M}_z$ enforces $D_{\parallel}=0$. Consequently, exchange frustration alone is insufficient to stabilize non-collinear states in these systems, leaving all four $\mathrm{Cr_2X_2}$ materials in the AM state, with either IP or OOP spin orientations depending on their magnetic anisotropy.
In contrast, breaking $\mathcal{M}_z$ in the Janus $\mathrm{Cr_2XY}$ materials activates a finite $D_{\parallel}$, giving rise to diverse non-collinear chiral spin structures. 

Specifically, $\mathrm{Cr_2SO}$, which combines strong frustration with a negligible $D_{\parallel}$ of $0.01$~meV, stabilizes high-$Q$ bimeron clusters. $\mathrm{Cr_2SeO}$ hosts similar high-$Q$ bimeron clusters, with a slight morphological deviation arising from its larger $D_{\parallel}$ of $0.11$~meV. 
Owing to its rather weak IP anisotropy ($K=-6.19~\mu\mathrm{eV/Cr}$, the smallest $|K|$ among the Janus monolayers studied here), the moderate $D_{\parallel}$ of 0.29 meV in $\mathrm{Cr_2SeS}$ is sufficient to tilt the spins out of plane, producing  isolated skyrmions each with $Q = 1$ and hence a total charge of high $Q$.
The contrast with $\mathrm{Cr_2TeO}$, which has a much stronger in-plane anisotropy ($K=-111.08~\mu\mathrm{eV/Cr}$) and retains high-$Q$ bimeron chains despite its larger $D_{\parallel}$ of $0.50$~meV, shows that the in-plane anisotropy competes with $D_{\parallel}$ in selecting between bimeron and skyrmion topologies. 
$\mathrm{Cr_2TeS}$, with the largest $D_{\parallel}$ of $1.00$~meV and OOP anisotropy, develops labyrinthine stripe domains (Lbr). 
Finally, $\mathrm{Cr_2TeSe}$, which exhibits OOP magnetic anisotropy, stabilizes isolated $Q=1$ skyrmions, highlighting the combined effects of $D_{\parallel}$, exchange frustration, and magnetic anisotropy in the generation of various topological spin textures.

To uncover the energy contributions stabilizing these bimeron and skyrmion states, as depicted in Fig.~\ref{fig_6}(d), we calculated the minimum energy paths for topological collapse into the uniform AM ground state using the GNEB method for two representative materials, $\mathrm{Cr_2SeO}$ and $\mathrm{Cr_2TeSe}$, which represent the frustration-dominated and DMI-dominated regimes, respectively.
The energy profiles reveal distinct barrier magnitudes and underlying stabilization mechanisms. For the $Q=3$ bimeron in frustration-dominated $\mathrm{Cr_2SeO}$, the saddle-point energy barrier $h$ is $12.14$~meV and is governed predominantly by the Heisenberg exchange energy, indicating that exchange frustration provides the main energetic protection against collapse. In contrast, the $Q=1$ skyrmion in DMI-dominated $\mathrm{Cr_2TeSe}$ exhibits a much higher barrier of $131.53$~meV, with the dominant contribution arising from $D_{\parallel}$. These results demonstrate that topological spin textures can be stabilized through distinct energetic mechanisms, with exchange frustration and $D_{\parallel}$ providing the dominant protection in the respective regimes.

Taken together, these results classify the monolayers into three regimes according to their topological character and dominant stabilization mechanism: topologically trivial (the four centrosymmetric $\mathrm{Cr_2X_2}$ monolayers), frustration-dominated ($\mathrm{Cr_2SO}$, $\mathrm{Cr_2SeO}$, and $\mathrm{Cr_2TeO}$, hosting high-$Q$ bimeron clusters and chains, together with $\mathrm{Cr_2SeS}$, where two isolated $Q=1$ skyrmions appear as a result of its weak in-plane anisotropy), and DMI-dominated ($\mathrm{Cr_2TeS}$ and $\mathrm{Cr_2TeSe}$, hosting Lbr and isolated $Q=1$ skyrmion states). In particular, the distinct behaviors of $\mathrm{Cr_2SeS}$ and $\mathrm{Cr_2TeS}$ suggest that frustration and $D_{\parallel}$ do not simply add up, which motivates the systematic parameter-space study presented in Sec.~\ref{subsec:IIID} D.

\subsection*{D. Interplay between magnetic frustration and DMI \label{subsec:IIID}}

To clarify the cooperative and competitive interplay between magnetic frustration and in-plane DMI, as illustrated in Fig.~\ref{fig_7}, we  scan the $\eta_1-D_{\parallel}$ and $\eta_2-D_{\parallel}$ parameter spaces for two representative prototypes, namely the frustration-dominated $\mathrm{Cr_2SeO}$ and the DMI-dominated $\mathrm{Cr_2TeSe}$. 
For each prototype, $\eta_1$, $\eta_2$, and $D_{\parallel}$ are independently varied from zero to twice their respective DFT-derived values.
To account for the possible existence of high-$Q$ states, all simulations are initialized with $Q=3$ skyrmion and bimeron configurations for the OOP and IP anisotropy cases, respectively. When a high-$Q$ state is not stabilized under a given set of parameters, the same initial configuration can relax into a corresponding lower-$Q$ state, allowing both high- and low-$Q$ phases to be captured within the parameter scans.

The energy difference relative to the collinear AM state $E-E_{\mathrm{AM}}$ maps the phase boundaries, as shown in Figs.~\ref{fig_7}(a,b) for $\mathrm{Cr_2SeO}$ and Figs.~\ref{fig_7}(g,h) for $\mathrm{Cr_2TeSe}$. 
For $\mathrm{Cr_2SeO}$, the boundary separating the AM phase from the topological bimeron phase depends primarily on the frustration parameters $\eta_1$ and $\eta_2$, exhibiting an almost flat contour that is nearly independent of $D_{\parallel}$. This behavior indicates that exchange frustration is the primary driving mechanism for the formation of high-$Q$ topological textures in this regime. 
For $\mathrm{Cr_2TeSe}$, the skyrmion phase can be reached through two distinct pathways, either by increasing frustration or by increasing $D_{\parallel}$. While increasing frustration produces a frustration-controlled boundary similar to that in $\mathrm{Cr_2SeO}$, increasing $D_{\parallel}$ reveals a clear signature of synergy: stronger frustration progressively lowers the critical $D_{\parallel}$ required to stabilize topological spin textures. Further increasing $D_{\parallel}$ drives the system into a Lbr stripe state, whose phase boundary also shifts toward lower $D_{\parallel}$ with increasing frustration. Within this Lbr regime, stronger frustration raises the topological charge $Q$ to values above $1$ (see Sec.~S5 in the SM~\cite{supplmat}), which already reveals the competition between frustration and $D_{\parallel}$ in setting the topological charge.

To quantify the role of $D_{\parallel}$ in shaping the topological spin textures, we introduce a normalized chiral correlation $\chi = \frac{1}{N} \sum_{\langle i,j\rangle} \hat{\mathbf{D}}_{ij}^{\parallel} \cdot (\mathbf{S}_i \times \mathbf{S}_j)$ which quantifies the global chiral ordering, with the unit vector along the in-plane DMI $\hat{\mathbf{D}}_{ij}^{\parallel}$, and the number of nearest-neighbor atom pairs $N$~\cite{z6k3-1zvy}. 
The $\chi$ maps in Figs.~\ref{fig_7}(c-d) and \ref{fig_7}(i-j) further resolve the sub-phases within the topological regions of $\mathrm{Cr_2SeO}$ and $\mathrm{Cr_2TeSe}$, respectively. 
For frustration-dominated $\mathrm{Cr_2SeO}$, the topological region is divided into high-$Q$ bimeron clusters and high-$Q$ bimeron chains, confirming that the morphological transition from cluster to chain is directly driven by $D_\parallel$. Trajectory analyses along the dotted paths further illustrate this evolution. At fixed $D_{\parallel}=0.17$~meV, increasing $\eta_1$ drives the texture from a bimeron cluster to an elongated cluster and eventually to a bimeron chain as $\eta_1$ increases from $0.14$ to $0.32$ [Fig.~\ref{fig_7}(e)]. 
Similarly, at fixed $\eta_2=0.09$, increasing $D_{\parallel}$ from $0$ to $0.21$~meV induces a transition from a bimeron cluster to a bimeron chain [Fig.~\ref{fig_7}(f)]. These results show that although $D_{\parallel}$ has little effect on the phase boundary of the high-$Q$ states, it still cooperates strongly with exchange frustration in shaping their magnetic structures.

For DMI-dominated $\mathrm{Cr_2TeSe}$, the topological region is divided into three sub-phases, namely high-$Q$ skyrmions, multiple $Q=1$ skyrmions, and isolated $Q=1$ skyrmions, demonstrating that $D_{\parallel}$ sensitively controls the topological charge $Q$. We further analyze the two trajectories marked in Fig.~\ref{fig_7}(j). Along the path at fixed $\eta_2=0.06$, increasing $D_{\parallel}$ drives the $Q=3$ skyrmion from a circular to an elliptical and eventually a peanut-shaped morphology, before it splits into multiple $Q=1$ skyrmions [Fig.~\ref{fig_7}(k)]. Conversely, along the path at fixed $D_{\parallel}=0.45$~meV, increasing $\eta_2$ induces a transition from a isolated $Q=1$ skyrmion to two $Q=1$ skyrmions ($Q=2$ in total), and finally to a high-$Q$  skyrmion with $Q=3$ [Fig.~\ref{fig_7}(l)]. 
The two trajectories thus demonstrate the competitive effect directly, since increasing $D_{\parallel}$ at fixed frustration reduces $Q$, whereas increasing frustration at fixed $D_{\parallel}$ raises it. 

In addition to the frustration-dominated and DMI-dominated regions, the parameter scans also reveal a broad intermediate regime, where the topological textures exhibit mixed characteristics. For instance, in $\mathrm{Cr_2SeO}$, elongated bimeron clusters appear between compact clusters and chains; in $\mathrm{Cr_2TeSe}$, elliptical and peanut-shaped $Q=3$ skyrmions serve as transitional configurations between high-$Q$ and multiple $Q=1$ states. These intermediate states underscore the continuous tuning effect of the balance between frustration and $D_{\parallel}$. Therefore, we identify a competitive effect, in which frustration favors high-$Q$ configurations, whereas $D_{\parallel}$ favors low-$Q$ states. 
Together with the synergistic effect identified above, where frustration lowers the critical $D_{\parallel}$, these two aspects constitute a dual mechanism governing the rich topological phenomenology observed in this system. This dual mechanism is consistent with the GNEB results in Sec.~\ref{subsec:IIIC} C, where the collapse barrier of the high-$Q$ bimeron is dominated by exchange frustration while that of the $Q=1$ skyrmion is dominated by $D_{\parallel}$.

\section{Conclusion}

In summary, our results show that the zero-field spin textures of monolayer altermagnets $\mathrm{Cr_2X_2}$ and Janus $\mathrm{Cr_2XY}$ ($\mathrm{X, Y}=\mathrm{O, S, Se, Te}$, and $\mathrm{X \neq Y}$) are set by exchange frustration, the in-plane DMI, and the magnetic anisotropy. Within a combined first-principles and atomistic spin-simulation scheme, we find that all effective exchange couplings are AFM, so frustration is of bond type and originates from same-sign rather than conventional ferro-antiferro competition, whereas the DMI appears only in the Janus configuration once the out-of-plane mirror symmetry is broken, following the Fert-Levy mechanism. Under zero external field, this interplay stabilizes topological spin textures on the compensated AM lattice, ranging from high-$Q$ bimeron clusters and chains in IP systems to robust $Q=1$ skyrmions in OOP ones. Notably, the GNEB calculations and parameter-space scans show that frustration and DMI both cooperate and compete, since frustration softens the collinear AM state and reduces the DMI strength needed to reach a topological state, while the two mechanisms set the charge in opposite directions, with frustration favoring high-$Q$ configurations and DMI favoring $Q=1$ states. The collapse barriers reflect the same duality, because the barrier of the high-$Q$ bimeron is governed by exchange frustration whereas that of the $Q=1$ skyrmion is governed by DMI. These results establish the $\mathrm{Cr_2XY}$ family as a versatile AM platform for exploring frustration-DMI interplay and provide a framework for designing field-free topological spin textures in compensated magnetic systems.

 {\textbf{Acknowledgments.}} This work was supported by National Natural Science Foundation of China (Grant No.11804301), the Natural Science Foundation of Zhejiang Province (Grant No. LMS25A040001), the Funds of the Natural Science Foundation of Hangzhou (Grant No. 2025SZRJJ0830), and Science Foundation of Zhejiang Sci-Tech University (Grant No. 26062265-Y).

 {\textbf{Data Availability.}} The datasets generated and analyzed in this study are not publicly available due to their large size and the associated cost of storage and hosting. The data are available from the corresponding author upon reasonable request.

\bibliography{References}

\end{document}